\documentclass[journal]{IEEEtran}
\IEEEoverridecommandlockouts
\usepackage{graphicx}
\usepackage{caption}
\usepackage{amsmath,amsthm,amsfonts,amssymb}
\usepackage{cite}
\usepackage{bm}
\usepackage{bbm}
\usepackage{url}
\usepackage{array}
\usepackage{color,soul}
\usepackage{multirow}
\usepackage{booktabs}
\usepackage[table,xcdraw]{xcolor}
\usepackage{enumitem}
\theoremstyle{plain}

\usepackage{subcaption}

\usepackage[english]{babel}
\usepackage{algorithm}
\usepackage[noend]{algpseudocode}

\usepackage{caption}
\usepackage[utf8]{inputenc}  
\usepackage{CJKutf8}         
\usepackage{comment}         
\usepackage{algorithm}       
\usepackage{algpseudocode}   
\usepackage{stfloats}
\allowdisplaybreaks[4]
\usepackage{graphicx} 
\usepackage{cases}   
\usepackage{xcolor}   
\begin{document}
\title{When Feedback Empowers the Uplink: Integrating Adaptive Coding with Wireless Power Transfer}
\author{Zijian Yang, Yulin Shao, Shaodan Ma
	\thanks{Z. Yang and S. Ma are with the State Key Laboratory of Internet of Things for Smart City and the Department of Electrical and Computer Engineering, University of Macau (emails:\{yc37408,shaodanma\}@um.edu.mo).}
	\thanks{Y. Shao is with the Department of Electrical and Electronic Engineering, University of Hong Kong (email:ylshao@hku.hk).}
}
\maketitle

\begin{abstract}
Energy consumption and device lifetime are critical concerns for battery-constrained IoT devices. This paper introduces the Feedback-Aided Coding and Energy Transfer (FACET) framework, which synergistically combines adaptive feedback channel coding with wireless power transfer. FACET leverages the saturation effect of feedback coding, where increasing downlink power yields diminishing returns, to design a dual-purpose feedback mechanism that simultaneously guides uplink coding and replenishes device energy. We characterize the inherent tradeoff between feedback precision and harvested power, and formulate a fairness-constrained min-max optimization problem to minimize worst-case net energy consumption. An efficient algorithm based on alternating optimization and Lagrangian duality is developed, with each subproblem admitting a closed-form solution. Simulations show that FACET nearly triples device lifetime compared to conventional feedback coding architectures, and remains robust across a wide range of power regimes. These results suggest that FACET not only improves communication efficiency but also redefines the role of feedback in energy-constrained IoT systems.
\end{abstract}


\section{Introduction}
The Internet of Things (IoT) has emerged as a cornerstone of modern technological ecosystems, enabling seamless connectivity across applications ranging from smart city infrastructures to industrial automation and precision healthcare \cite{IoT1,nbiot,shao2021federated}. Despite this rapid expansion, most IoT devices are battery-powered and often deployed in environments where frequent recharging or battery replacement is impractical \cite{nbiot,LoRa}.  Conventional energy-saving approaches, including Power Saving Mode (PSM) and enhanced Discontinuous Reception (eDRX), can indeed extend device lifetime by reducing idle listening or prolonging sleep intervals \cite{LPWA}. However, these strategies only postpone the inevitable depletion of battery resources. This stems from a more fundamental inefficiency: unidirectional, transmitter-dominated communication. For instance, in Narrowband IoT (NB-IoT), devices repetitively transmit data packets up to 128 times to overcome extreme path loss, draining limited battery reserves with each attempt.

In response to these challenges, feedback-enhanced IoT communication pioneers a new paradigm that redefines the relationship between IoT devices and their access points (APs) \cite{DEEP_IoT,Polyanskiy}. Rather than forcing IoT devices to repeatedly transmit and blindly hope for reception, feedback-enhanced IoT grants the AP a more proactive role by providing real-time decoding feedback, enabling each transmitter to dynamically refine its uplink coding strategy on the fly. Crucially, this refinement is propelled by advanced feedback channel codes \cite{Shannon,deepcode,attentioncode,ALLyouneed}, which facilitate more granular and timely decoding information. 
By thus shifting the energy burden to the AP, whose power supply is far more robust, feedback-coding-aided IoT trades the AP's abundant energy for the IoT device's limited battery power, culminating in a ``listen more, transmit less'' philosophy \cite{DEEP_IoT}. In this way, strategic downlink feedback empowers the uplink by eliminating wasteful uplink retransmissions, achieving significant reductions in energy consumption.

While feedback-coding-aided IoT effectively shift the energy burden from power-constrained IoT devices to energy-abundant APs, a critical characteristic of feedback channel codes is the saturation effect: as downlink signal quality improves, the feedback coding gain diminishes and ultimately converges to a theoretical limit \cite{Polyanskiy}. This diminishing return reveals a fundamental inefficiency that any surplus feedback power becomes underutilized once feedback quality is ensured. Motivated by this insight, we pose a critical question: {\it Could the feedback power also be harnessed by the IoT device via
wireless power transfer (WPT)?}
This proposition reimagines the boundaries of feedback-aided systems, proposing a dual-purpose paradigm where downlink feedback not only guides
adaptive coding in the uplink but also physically replenishes the IoT devices they serve.

The main contributions of this paper are twofold:
\begin{itemize}[leftmargin=0.45cm]
    \item We put forth the Feedback-Aided Coding and Energy Transfer (FACET) framework, which synergistically integrates adaptive feedback channel coding with WPT. FACET exploits the saturation effect of feedback-aided adaptive coding, where the marginal benefit of increasing feedback power diminishes beyond a threshold, and introduces a dual-purpose feedback mechanism that significantly extends device operational longevity without compromising communication reliability.
    \item We characterize an inherent tradeoff in FACET between adaptive coding gains and harvested energy: increasing the fraction of downlink power allocated to WPT enhances energy replenishment but risks degrading feedback precision, which in turn elevates uplink transmit power. To balance this tradeoff, we formulate a fairness-constrained min-max optimization problem that minimizes the worst-case net energy consumption across all devices. Leveraging Lagrangian duality and alternating optimization, we develop a solution where each subproblem admits a closed-form solution, ensuring both computational efficiency and guaranteed convergence.
\end{itemize}

\section{System Model}\label{sec:II}
\begin{figure}[t]
    \centering
    \includegraphics[width=0.7\columnwidth]{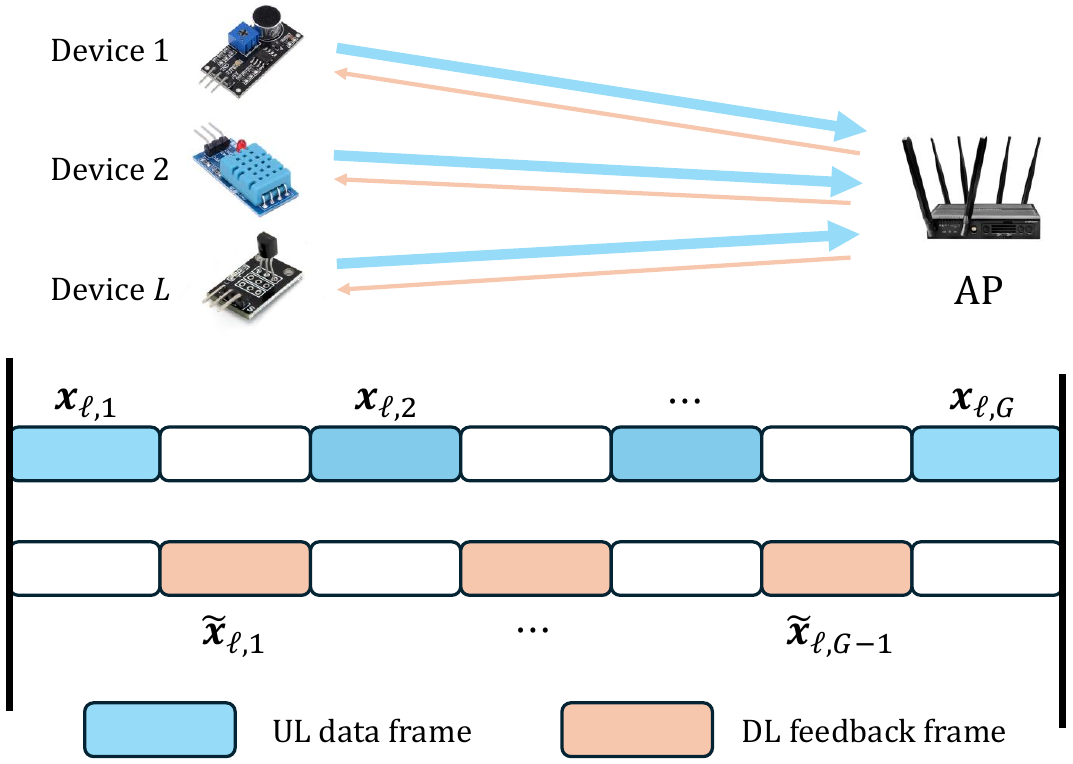}
    \caption{System model and frame structure of the feedback-aided IoT uplink, where uplink transmission and downlink feedback alternate in time for real-time adaptive coding.}
    \label{fig:1}
\end{figure}

We consider the uplink transmission from $L$ IoT devices to an AP, as illustrated in Fig.~\ref{fig:1}. 
The AP is equipped with $M$ antennas, while each IoT device has a single antenna.
To improve energy efficiency, the IoT devices operate with feedback-aided channel coding, where each device adapts its encoded symbols not only based on its source data bits but also on real-time decoding feedback from the AP. This dynamic adaptation allows devices to transmit with reduced power while still ensuring reliable decoding at the AP.

Denote by $\bm{b}_\ell \in \{0, 1\}^{K \times 1}$ the data to be transmitted from the $\ell$-th IoT device to the AP.
With feedback channel coding, the uplink transmission is divided into $G$ frames, each consisting of $Q$ complex coded symbols, denoted by $\bm{x}_{\ell,g} \in \mathbb{C}^{Q \times 1}$ for $g = 1, 2, \ldots, G$. Right after the $g$-th uplink frame, the AP feeds back a downlink frame $\widetilde{\bm{x}}_{\ell,g} \in \mathbb{C}^{Q \times 1}$, which informs and guides the construction of the subsequent uplink frame $\bm{x}_{\ell,g+1}$. 
After the completion of all $G$ uplink frames, the AP accumulates a total of $N = 2QG$ real channel symbols and attempts to decode the original bits, producing an estimate $\widehat{\bm{b}}$.
The entire process thus constitutes a feedback channel code \cite{Shannon,Polyanskiy}, characterized by $\mathcal{C}_f(K, N, \epsilon)$, where $\epsilon = \Pr(\bm{b} \neq \widehat{\bm{b}})$ represents the block error rate (BLER).

Following the design principles of NB-IoT, we consider the IoT cell operating in frequency-division duplexing mode with narrowband communication \cite{nbiot}. In this setting, uplink and downlink transmissions occupy separate frequency bands, and each transmission uses a single subcarrier to ensure high power spectral density, thereby achieving wide and deep coverage.

In the uplink, each IoT device is assigned a unique subcarrier. The signal received at the 
$m$-th antenna of the AP from device $\ell$ is given by
\begin{equation}
\bm{y}_{\ell,g}^{(m)} = \sqrt{d_\ell^{-\alpha} P_\ell} \, h_{\ell}^{(m)} \, \bm{x}_{\ell,g} + \bm{z}_{\ell,g}^{(m)},
\end{equation}
where
 $d_\ell$ is the distance between the AP and the $\ell$-th IoT device,
 $\alpha$ is the path loss exponent,
 $P_\ell$ is the transmit power of device $\ell$,
 $h_{\ell}^{(m)}$ is the channel coefficient for the $m$-th antenna and the $\ell$-th device,
 $\bm{z}_{\ell,g}^{(m)} \sim \mathcal{CN}(\bm{0}, \sigma^2 \bm{I}_Q)$ is the additive white Gaussian noise (AWGN) vector,
 and the power of $\bm{x}_{\ell,g}$ is normalized to 1.
 
Applying maximum ratio combining, the resulting signal-to-noise ratio (SNR) at the AP for device $\ell$ is $\eta_\ell = \frac{P_\ell d_\ell^{-\alpha}}{\sigma^2}
\sum_{m=1}^{M} \left| h_{\ell}^{(m)} \right|^2$.
We also write SNR in decibels as $\eta_{\ell,\text{dB}}$.

In the downlink, subcarrier allocation is dynamic and depends on resource availability. Let $S$ denote the total number of available subcarriers, and define $\delta_{s,\ell} \in \{0, 1\}$ as the binary indicator for subcarrier allocation:
\begin{equation*}
\delta_{s,\ell} =
\begin{cases}
1, & \text{if subcarrier $s$ is assigned to device $\ell$}, \\
0, & \text{otherwise}.
\end{cases}
\end{equation*}
To maintain orthogonality and exclusive allocation, these two constraints hold: $\sum_{\ell=1}^{L} \delta_{s,\ell} \leq 1$ and $\sum_{s=1}^{S} \delta_{s,\ell} = 1$.

The signal received by device $\ell$ is then
\begin{equation}
\widetilde{\bm{y}}_{\ell,g} = \sum_{s=1}^{S} \delta_{s,\ell} \sqrt{d_\ell^{-\alpha} \widetilde{p}_{s}} \, \widetilde{\bm{h}}_{s,\ell}^{\top} \bm{w}_{s} \, \widetilde{\bm{x}}_{\ell,g} + \widetilde{\bm{z}}_{\ell,g},
\end{equation}
where $\bm{w}_s \in \mathbb{C}^{M \times 1}$ is the precoding vector for subcarrier $s$, and 
$
\widetilde{\bm{h}}_{s,\ell} \triangleq \left[\widetilde{h}_{s,\ell}^{(1)}, \widetilde{h}_{s,\ell}^{(2)}, \dots, \widetilde{h}_{s,\ell}^{(M)}\right]^\top
$ is the downlink channel vector from the AP to device $\ell$ on subcarrier $s$. Since the system operates in FDD mode, the downlink and uplink channels are independently modeled. 
The transmit power allocated to subcarrier $s$ is denoted by $\widetilde{p}_s$, subject to the total feedback power constraint $\sum_{s=1}^{S} \widetilde{p}_s \leq \widetilde{P}_{\text{total}}$.
The AWGN vector $\widetilde{\bm{z}}_{\ell,g} \sim \mathcal{CN}(\bm{0}, \widetilde{\sigma}^2 \bm{I}_Q)$.

Under maximum ratio transmission precoding, the downlink SNR for device $\ell$ is given by
\begin{equation}\label{eq:DLSNR}
\widetilde{\eta}_{\ell} = \sum_{s=1}^{S}\frac{\delta_{s,\ell} \, \widetilde{p}_s d_\ell^{-\alpha} \| \widetilde{\bm{h}}_{s,\ell} \|^2}{\widetilde{\sigma}^2}.
\end{equation}

With real-time decoding feedback from the AP, each IoT device can adapt its channel coding strategy in subsequent uplink frames, thereby improving coding efficiency. For a given feedback code $C_f(K, N, \epsilon)$ and a target BLER $\epsilon^*$, the critical uplink SNR $\eta_{\ell,\text{dB}}$ required to meet $\epsilon^*$ depends on the feedback SNR $\widetilde{\eta}_{\ell,\text{dB}}$. To characterize the power reduction capability of feedback-aided adaptive channel coding, we consider the feedback codes in \cite{DEEP_IoT}, where the critical uplink SNR to achieve $\epsilon^*$ is characterized as
\begin{equation}\label{eq:deepiot}
\eta_{\ell,\text{dB}} = u_0(\epsilon^*) + \frac{1}{\exp\left\{ u_1(\epsilon^*) + u_2(\epsilon^*) \widetilde{\eta}_{\ell,\text{dB}}  \right\} + u_3(\epsilon^*)},
\end{equation}
This expression reveals that the critical uplink SNR decreases exponentially with increasing feedback SNR. As the feedback quality improves, the AP can successfully decode uplink transmissions even at lower transmit power, thanks to better-informed adaptive coding. This relationship highlights the core advantage of feedback coding: it shifts the energy burden from battery-constrained IoT devices to the energy-abundant AP.

However, \eqref{eq:deepiot} also highlights a saturation effect: as the feedback SNR continues to improve, the resulting reduction in required uplink SNR becomes increasingly marginal, eventually approaching the theoretical lower bound 
$u_0(\epsilon^*)$. This diminishing return is consistent with classical information-theoretic limits \cite{Polyanskiy}: while feedback can significantly enhance reliability and reduce power, its capacity benefit is fundamentally bounded. This insight motivates the integration of WPT into the feedback framework, allowing excess downlink power to be redirected toward energy harvesting once the feedback quality has reached a sufficient level.
\section{Feedback-Aided Coding and Energy Transfer}
Section~\ref{sec:II} establishes the feedback-aided communication model in which real-time decoding feedback from the AP allows IoT devices to dynamically adapt their uplink channel coding strategies, thereby reducing transmission energy. While these downlink feedback signals primarily serve an informational role, they also inherently carry radio frequency (RF) energy. This observation motivates a new opportunity: can the feedback signals be used to wirelessly power the devices they serve?
This section, we introduce our FACET framework that jointly exploits the downlink feedback signal for both feedback coding and energy harvesting.

\subsection{The Fundamental Tradeoff of FACET}
FACET extends the conventional feedback-aided architecture by integrating an RF energy harvesting module into each IoT device. While each device only decodes feedback on its allocated subcarrier, it can harvest energy from all downlink transmissions across the frequency band. The total received RF power at device $\ell$ is given by
\begin{equation}
\widetilde{P}_\ell = \sum_{s=1}^{S}\sum_{j=1}^{L} \widetilde{p}_s d_\ell^{-\alpha} \delta_{s,j}\left\| \widetilde{\bm{h}}_{s,\ell}^H \right\|^2.
\end{equation}

To support simultaneous feedback decoding and energy harvesting, each device employs a power-splitting mechanism: a portion $\rho_\ell \in [0,1]$ of the received power is directed to the energy harvester \cite{zhang2018wireless,clerckx2018fundamentals}, and the remaining $1 - \rho_\ell$ is used for feedback decoding.
This introduces a fundamental tradeoff: increasing $\rho_\ell$ improves the harvested energy but reduces the signal power available for decoding, potentially degrading the feedback quality and increasing the uplink transmit power required to maintain reliability. Conversely, reducing $\rho_\ell$ enhances decoding quality and uplink coding efficiency but limits the harvested energy. As a result, feedback quality and harvested energy are inherently coupled via $\rho_\ell$, and FACET must navigate this tradeoff intelligently.

To balance the tradeoff, we formulate a joint optimization problem that seeks to minimize the maximum net energy consumption across all devices. Specifically, we define the net energy as the difference between the energy consumption of uplink transmission and the harvested energy at each device. By minimizing the worst-case net energy, the system ensures max-min fairness, providing uniform energy efficiency regardless of device location or channel condition. Mathematically, the optimization objective can be written as
\begin{subequations} \label{objective1}
\begin{align}
&\underset{\{\widetilde{p}_s, \delta_{s,\ell}, \rho_\ell\}}{\text{min}} \quad  \max_{\ell \in \{1, \dots, L\}} (P_\ell -  \widetilde{E}_\ell)\\
 &{ ~\text {s.t.}~}
    \sum_{s=1}^{S} \widetilde{p}_s \leq \widetilde{P}_{\text{total}},~  
    \widetilde{p}_s \geq 0,~\forall s,~  
    0 \leq \rho_\ell \leq 1,~ \forall \ell,  \label{objective1_constraint_b} \\
    &\delta_{s,\ell} \in \{0, 1\}, \forall \ell, s,~  
    \sum_{\ell=1}^{L} \delta_{s,\ell} \leq 1,  \forall s, ~
    \sum_{s=1}^{S} \delta_{s,\ell} \!=\! 1, \forall \ell.  \label{objective1_constraint_c}
\end{align}
\end{subequations}

In \eqref{objective1}, the harvested energy $\widetilde{E}_\ell = \rho_\ell \kappa \widetilde{P}_\ell$, where $\kappa\in(0,1)$ denotes the harvesting efficiency factor.
Moreover, since only a portion $1 - \rho_\ell$ of the received signal is utilized for feedback decoding, the effective downlink SNR in \eqref{eq:DLSNR} should be scaled accordingly. Therefore, we rewrite it as
$\widetilde{\eta}_{\ell} = \sum_{s=1}^{S}\frac{\delta_{s,\ell} \, \widetilde{p}_s (1 - \rho_\ell) d_\ell^{-\alpha} \| \widetilde{\bm{h}}_{s,\ell} \|^2}{\widetilde{\sigma}^2}$.

\subsection{Solving the Fundamental Tradeoff}
The resource allocation problem in FACET presents a mixed-integer non-convex optimization challenge due to the coupling between binary subcarrier assignments and continuous power-related variables. To efficiently solve this, we develop a two-stage solution strategy that separates the discrete and continuous components of the problem.

In the first stage, we address the subcarrier assignment through a Hungarian algorithm-based method \cite{hungarian_algorithm}. The objective is to maximize the total weighted channel gain across all devices, while implicitly addressing fairness by prioritizing users with higher channel quality. The subcarrier allocation subproblem can be written as
\begin{subequations}
\label{subcarrier}
\begin{align}
    &\underset{\delta_{s,\ell}}{\text{max}} \sum_{s=1}^{S} \sum_{\ell=1}^{L} 
    \delta_{s,\ell} d^{-\alpha_\ell} \|\widetilde{\bm{h}}_{s,\ell}\|^2, \quad
  ~\text {s.t.}~ \eqref{objective1_constraint_c}. 
  \end{align}
\end{subequations}
This formulation constitutes a classical one-to-one assignment problem, where each IoT device is assigned a unique subcarrier to maximize the total channel gain across all users. The Hungarian algorithm efficiently solves this problem with polynomial complexity, yielding a globally optimal matching. While it does not guarantee that every device receives its individually strongest channel, it ensures the overall sum of channel gains is maximized, thereby achieving a balance between global efficiency and individual performance.

In the second stage, given the subcarrier assignments, we solve for the optimal power allocations and power-splitting ratios. This is accomplished by reformulating the original problem into a set of convex subproblems, which are then solved using alternating optimization combined with Lagrangian duality. This decomposition yields closed-form solutions for each variable block, ensuring both computational efficiency and provable convergence to a local optimum.

To start with, we introduce an auxiliary variable $t_\ell$ to represent the critical uplink SNR for each device. This variable replaces the nonlinear expression $\eta_{\ell,\text{dB}}$, allowing us to decouple the downlink feedback constraint into a more tractable form. For notational simplicity, we define three constants that encapsulate key propagation and noise factors: $U^{(0)}_\ell\triangleq\frac{\sigma^2}{d_\ell^{-\alpha}\|\bm{h}_\ell\|^2}$, $U^{(1)}_{s,\ell}\triangleq\kappa d_\ell^{-\alpha}\|\widetilde{\bm{h}}_{s,\ell}\|^2$, and $U^{(2)}_{s,\ell} \triangleq \frac{d_\ell^{-\alpha} \| \widetilde{\bm{h}}_{s,\ell} \|^2}{\widetilde{\sigma}^2}$.

The optimization problem for power allocations and power-splitting ratios can be transformed into 
\begin{subequations}\label{equation_t}
\begin{align}
\underset{\{\widetilde{p}_s, \rho_\ell, t_\ell\}}\min  \max_{\ell \in \{1, \dots, L\}} &\sum_{s=1}^{S}\sum_{j=1}^{L} \left( U^{(0)}_\ell 10^{\frac{t_\ell}{10}} - U^{(1)}_{s,\ell} \rho_\ell \widetilde{p}_{s} \delta_{s,j} \right)\\
~\text {s.t.}~\eqref{objective1_constraint_b},~
    &\eta_{\ell,\text{dB}} \leq t_\ell, \quad \forall \ell.
    \label{constraint_eta_t}
\end{align}
\end{subequations}

Substituting \eqref{eq:deepiot} into \eqref{constraint_eta_t},  we have
\begin{equation}
u_0 + \frac{1}{\exp\left(u_1 + u_2 \widetilde{\eta}_{\ell,\text{dB}}\right) +u_3} \leq t_\ell
.
\end{equation}
We can transform \eqref{constraint_eta_t} into logarithmic form since ${\exp (u_1\!+\!{u_2}{{\widetilde \eta }_{\ell,\text{dB} }})}\!=\!\exp(u_1)\allowbreak\big[ {\sum_{s=1}^{S}U^{(2)}_{s,\ell}\allowbreak(1 \!-\! {\rho _\ell })\allowbreak \delta _{s,\ell }\allowbreak\widetilde{p}_{s}}\big]^{\frac{{10{u_2}}}{{\ln 10}}}$, yielding
\begin{equation}\label{eq:constraint}
\frac{{\ln 10}}{{10{u_2}}}\left[ {\ln \left(\frac{1}{{t_\ell \!-\! {u_0}}} \!-\! {u_3}\right) \!-\! {u_1}} \right] \!\le\! \ln \left[ {\sum_{s=1}^{S}\!U^{(2)}_{s,\ell}(1 \!-\! \rho_\ell) \delta_{s,\ell} \widetilde{p}_{s}} \right].
\end{equation}

To promote fairness across users, we introduce a scalar variable 
$r$ representing the maximum net energy consumption, and reformulate the problem as
\begin{subequations}\label{equation_r}
\begin{align}
    &\underset{\{\widetilde{p}_s, \rho_\ell, t_\ell, r\}}{\text{min}} \quad  r \\
    &\text {s.t.}~ \eqref{eq:constraint}, \eqref{objective1_constraint_b}, \left( U^{(0)}_\ell 10^{\frac{t_\ell}{10}} - \rho_\ell\sum_{s=1}^{S}\sum_{j=1}^{L} U^{(1)}_{s,\ell}  \widetilde{p}_{s} \delta_{s,j} \right) \leq r,~\forall \ell, \notag
\end{align}
\end{subequations}

Compared with the original problem, the transformed problem introduces an additional set of $L$ constraints to ensure that the energy consumption of each device remains below $r$.

After the above manipulations, the original mixed-integer non-convex problem has been transformed into a convex optimization problem with respect to the continuous variables $\widetilde{p}_{s}$, $\rho_\ell$, and $t_\ell$. The convexity enables us to efficiently solve the problem using an alternating optimization approach.

\subsection{Closed-Form Solution via Lagrangian Duality}
To efficiently solve the reformulated convex problem, we adopt a dual decomposition approach combined with the Lagrangian duality method. The core idea is to break the joint optimization into two interrelated subproblems: one concerning the auxiliary variable $t_\ell$, and the other involving power allocation $\widetilde{p}_{s}$ and power splitting ratio $\rho_\ell$. These subproblems are alternately optimized using closed-form solutions derived from the Karush-Kuhn-Tucker (KKT) conditions, while the associated dual variables are updated via a subgradient method.

\subsubsection{Solutions for the auxiliary variables}
We begin with the subproblem concerning $t_{\ell}$, which influences the uplink power requirement through the feedback-dependent coding. The Lagrangian for this subproblem is given by
\begin{equation*}
\scalebox{0.76}{$\displaystyle   
\begin{aligned}
\mathcal{L}(r,\bm{t}, \bm{\gamma}, \bm{\nu}) = \, & r + \sum_{\ell=1}^{L}\gamma_\ell \left( U^{(0)}_\ell 10^{\frac{t_\ell}{10}} 
-\sum_{s = 1}^S\sum_{j=1}^{L} U^{(1)}_{s,\ell} \rho_\ell \widetilde{p}_{s}\delta_{s,j} - r \right) 
 + \sum_{\ell=1}^{L}\nu_\ell\\ &\left( t_\ell - \frac{1}{\exp(u_1) \cdot \left[\sum_{s=1}^{S}U^{(2)}_{s,\ell} (1 - \rho_\ell) 
\delta_{s,\ell} \widetilde{p}_{s} \right]^{\frac{10 u_2}{\ln 10}} + u_3} + u_0 \right).
\end{aligned}
$}   
\end{equation*}
Differentiating the Lagrangian with respect to $t_{\ell}$ and applying the KKT conditions yield a closed-form solution $
{t_\ell }^* = 10{\log _{10}}\frac{{-10{\nu _\ell }}}{{\gamma_\ell{U^{(0)}_\ell}\ln 10}}, \forall \ell.$

\subsubsection{Solutions for power allocation and power splitting ratios}
Next, we address the subproblem involving $\widetilde{p}_{s}$ and  $\rho_\ell$, which jointly determine the feedback quality and the harvested energy.
The corresponding Lagrangian function is
\begin{equation*}
\scalebox{0.78}{$\displaystyle 
    \begin{aligned}
    &{\cal L}(\bm{p},\bm{\delta},\bm{\rho},\bm{t},r,\bm{\lambda},\bm{\mu},\eta) 
    = r + \sum_{\ell=1}^{L} \lambda_\ell \bigg( U^{(0)}_\ell 10^{\frac{t_\ell}{10}} - \rho_\ell\sum_{s=1}^{S} U^{(1)}_{s,\ell} \sum_{j=1}^{L} \delta_{s,j} \widetilde{p}_{s} - r \bigg) \\
    &\quad + \sum_{\ell=1}^{L}\mu_\ell \bigg\{ \frac{\ln 10}{10 u_2} \bigg[ \ln \bigg( \frac{1}{t_\ell - u_0} - u_3 \bigg) - u_1 \bigg]  - \ln \bigg[ \sum_{s=1}^{S} U^{(2)}_{s,\ell} (1 - \rho_\ell) \delta_{s,\ell} \widetilde{p}_{s} \bigg] \bigg\} \\
    &\quad + \eta \bigg( \sum_{s = 1}^S  \widetilde{p}_{s} - P_{\text{total}} \bigg).
    \end{aligned}
$}
\end{equation*}

Applying the KKT conditions, the closed-form solutions for feedback power allocation and power splitting ratios are
$\widetilde{p}_{s}^*\!=\!(\sum_{j=1}^{L}\delta_{s,j})\left[\frac{\sum_{\ell=1}^{L}\mu_\ell }{{\eta - \sum_{\ell=1}^{L}\lambda_\ell {\rho _\ell }{U^{(1)}_{s,\ell}}\sum_{j=1}^{L}\delta_{s,j}}}\right]^+,\forall s$,
$\rho _\ell ^*\!=\!\left[1-\!\frac{\mu_\ell }{{  \lambda_\ell \sum_{s = 1}^S {U^{(1)}_{s,\ell}}{\widetilde{p}_{s}}\sum_{j=1}^{L}\delta_{s,j}}}\right]_0^1,\forall \ell$.

\subsubsection{Dual variable updates via subgradient method}
To complete the dual optimization, the Lagrange multipliers ($\gamma,\nu,\lambda,\mu,\eta$) are updated using subgradient steps. 
With proper initialization and step size selection, the alternating primal-dual updates converge to a stationary point. The proposed closed-form solution offers twofold advantages: it reveals the structure and dependencies of the optimal resource allocation, and it significantly reduces computational complexity compared to generic convex solvers. This makes it well-suited for practical deployment in large-scale IoT networks. 

\section{Performance Evaluation and Discussion}
\label{sec:results}


\begin{figure*}[t]
    \centering
    \begin{minipage}[t]{0.3\textwidth}
        \centering
        \includegraphics[width=\textwidth]{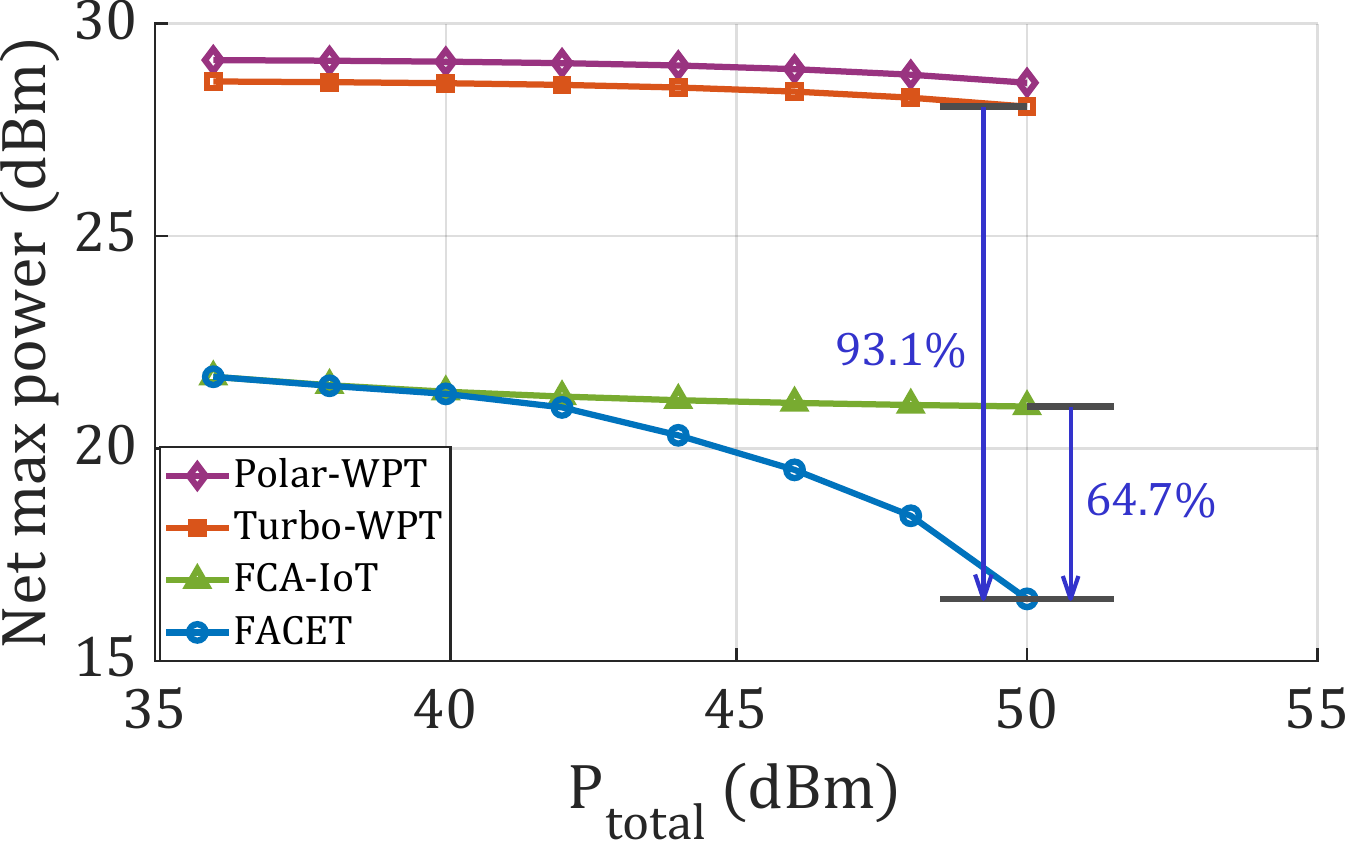}
      \subcaption{}
    \end{minipage}
    \hfill
    \begin{minipage}[t]{0.38\textwidth}
        \centering
        \includegraphics[width=\textwidth]{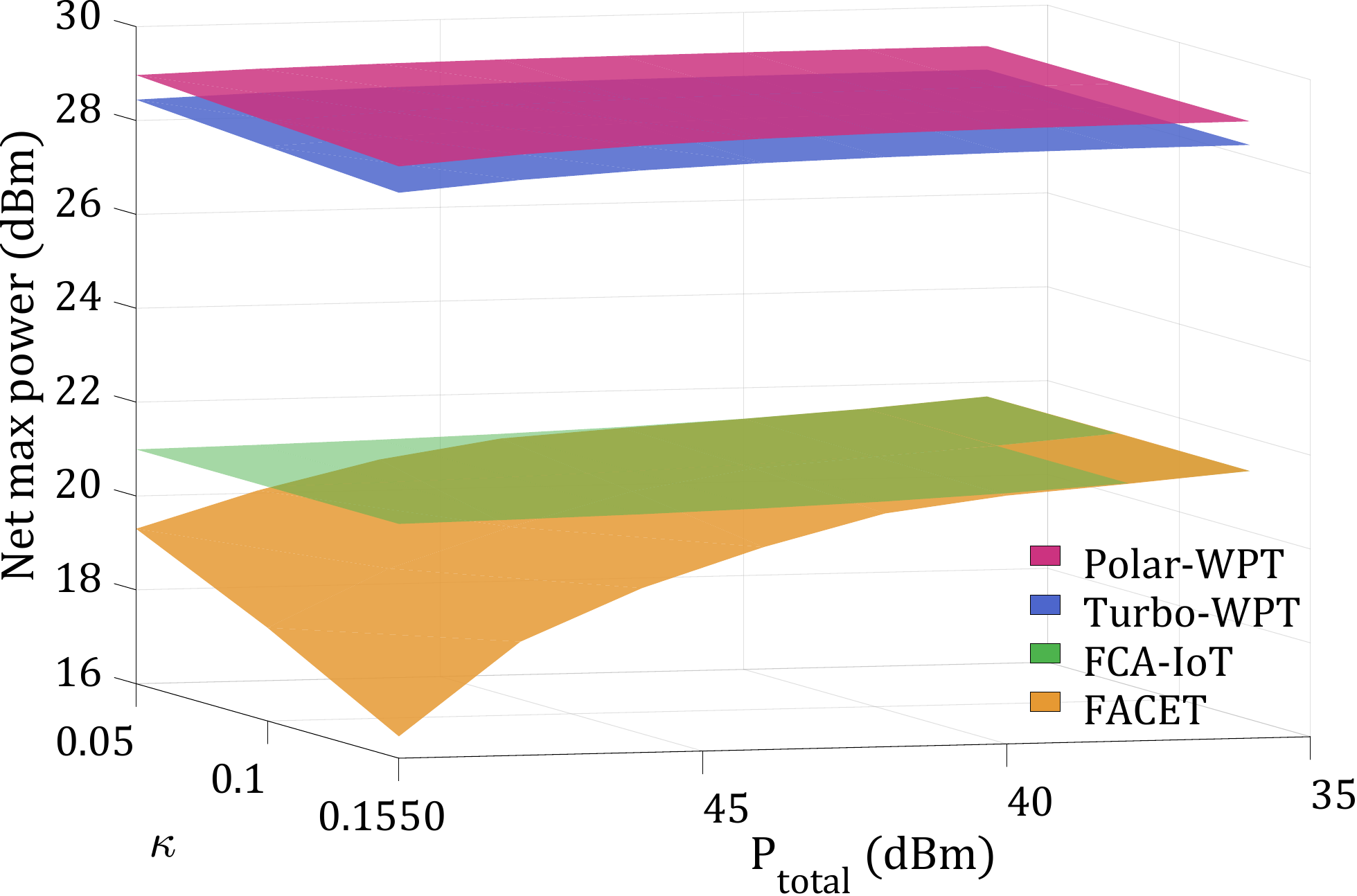}
       \subcaption{}

    \end{minipage}
    \hfill
    \begin{minipage}[t]{0.3\textwidth}
        \centering
        \includegraphics[width=\textwidth]{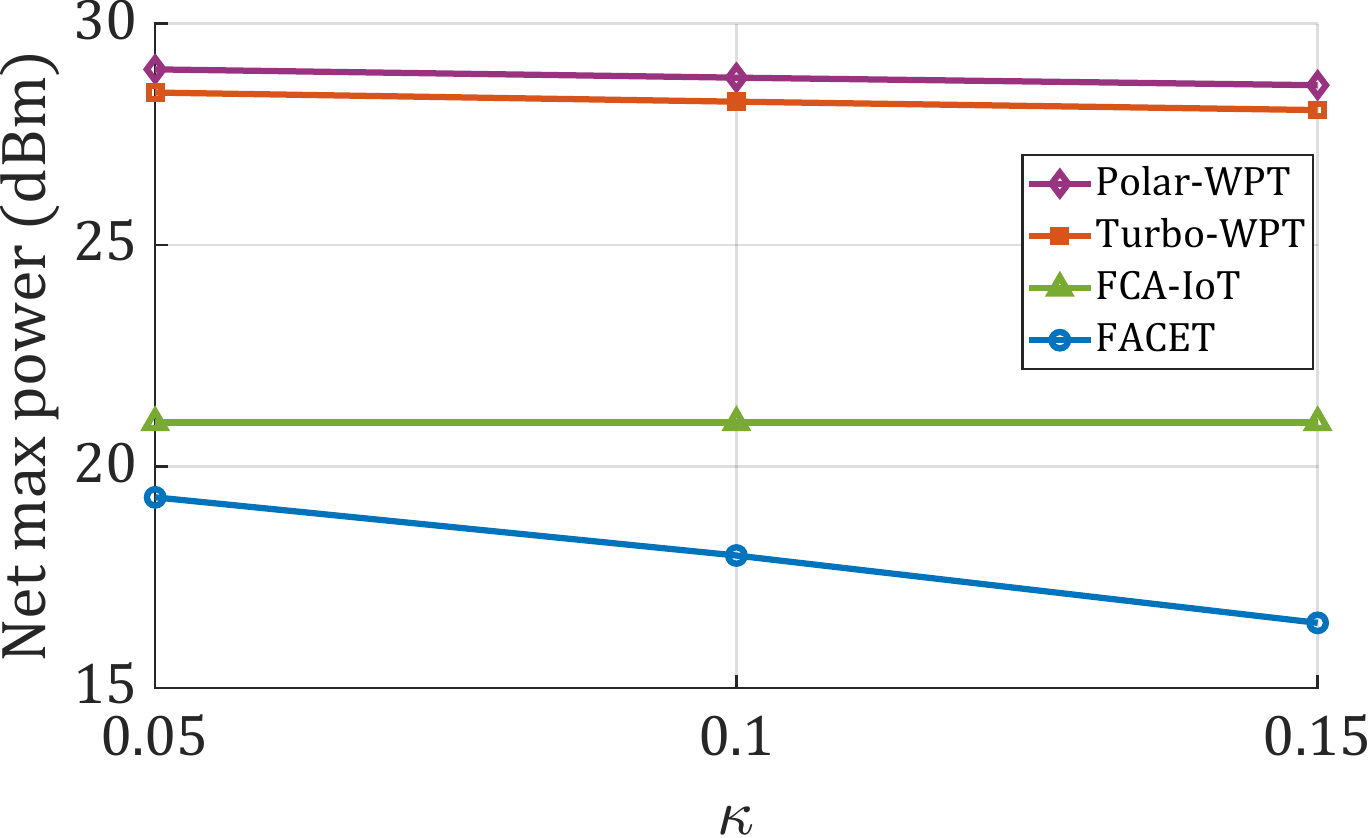}
       \subcaption{}

    \end{minipage}
    \caption{Net maximum power consumption under varying total transmit power and energy harvesting efficiency. The left and right subfigures show cross sections at $\kappa = 0.15$ and $P_{\text{total}} = 50\,\mathrm{dBm}$, respectively.}
    \label{fig:3d_map}
    
\end{figure*}

This section evaluates the performance of FACET benchmarked against three baselines: the Polar-coded WPT system (Polar-WPT), the Turbo-coded WPT system (Turbo-WPT), and the feedback-coding-aided IoT system without WPT (FCA-IoT). Simulations results are presented in Fig.~\ref{fig:3d_map}.

To begin with, we compare FACET with standard forward channel-coded systems with WPT. As shown, FACET outperforms both Turbo-WPT and Polar-WPT significantly. Compared to Turbo-WPT, FACET achieves a remarkable $93.1\%$ reduction in net maximum power consumption at $P_{\text{total}} = 50$dBm. This improvement translates to a $13.4$-fold increase in device lifetime under the same battery capacity. 
This performance gain stems from FACET's ability to dynamically split received power between energy harvesting and decoding, whereas Turbo-WPT and Polar-WPT use a fixed power-splitting ratio of $\rho = 1$, allocating all received energy to harvesting, as also reflected in Table~\ref{tab:comparison}.

In comparison to FCA-IoT, FACET exhibits a dual-regime behavior. 
In the low-power regime (e.g., $P_{\text{total}}\in[36, 42]$dBm), FACET performs on par with FCA-IoT, with a negligible performance gap. This aligns with our intuition because limited downlink feedback power forces the system to prioritize reliable decoding, leading to nearly full allocation of received energy to decoding (i.e., $\rho \to 0$). Consequently, both FACET and FCA-IoT emphasize feedback-aided channel coding over energy harvesting.
In contrast, in the high-power regime (e.g., $P_{\text{total}}\in[44, 50]$dBm), FACET achieves significant reduction in energy consumption over FCA-IoT. For quantitative comparison at $P_{\text{total}} = 50$dBm and $\kappa = 0.15$, Table~\ref{tab:comparison} summarizes key metrics across all schemes.
 For example, FACET achieves a $64.7\%$ reduction in net maximum power consumption compared to FCA-IoT, which corresponds to an estimated $1.83$-fold extension in device lifetime. In particular, as feedback power increases, the optimal $\rho$ shifts toward larger values, indicating that a significant portion of received power is diverted toward energy harvesting instead of decoding.

Further insights are provided by the right subfigure of Fig.~\ref{fig:3d_map}, which shows how the net maximum power consumption varies with the energy harvesting efficiency $\kappa$, under a fixed total feedback power of $50$dBm.
While FACET exhibits the largest reduction in power consumption on the dBm scale, Turbo-WPT and Polar-WPT achieve larger absolute reductions in the linear (Watt) domain thanks to their fixed $\rho=1$, making them more sensitive to changes in $\kappa$. In contrast, FACET employs a dynamically optimized power-splitting ratio, balancing between feedback decoding and WPT. Meanwhile, FCA-IoT does not harvest energy and thus remains unaffected by varying $\kappa$. These observations emphasize that FACET's advantage stems not from maximized energy harvesting, but from its ability to achieve a Pareto-optimal trade-off between energy harvesting and feedback-assisted decoding, ensuring consistent performance improvements across different energy efficiency conditions.

To summarize, the evaluations reveal two distinct performance regimes for FACET:
\begin{itemize}[leftmargin=0.4cm]
    \item \textbf{Coding-dominated regime.} When the feedback power $P_{\text{total}}$ is relatively small, energy harvesting plays a minimal role. FACET in this case behaves similarly to FCA-IoT, focusing primarily on improving the uplink coding efficiency through high-quality feedback. 
    \item \textbf{WPT-dominated regime.} As $P_{\text{total}}$ increases, surplus feedback power enables efficient energy harvesting, which substantially reduces the uplink transmission burden. The synergistic optimization of $\rho$ and downlink power allocation becomes essential in this regime. In contrast, static strategies like Polar-WPT and Turbo-WPT fail to exploit this surplus, leading to suboptimal performance.
\end{itemize}

\begin{table}[t]
    \centering
    \setlength{\tabcolsep}{2pt}
    \caption{Performances with $\kappa = 0.15$ and $P_{\text{total}} = 50$dBm.}
    \label{tab:comparison}
    \begin{tabular}{lcccc}
        \toprule
        \textbf{Metric} & \textbf{Polar-WPT}& \textbf{Turbo-WPT} & \textbf{FCA-IoT} & \textbf{FACET} \\
        \midrule
        Net Max Power (dBm) & 28.61 & 28.05 & 20.99 & \textbf{16.47} \\
        Net Max Power (Watt) & 0.73 & 0.64 & 0.13 & \textbf{0.04} \\ 
        Uplink Coding Gain & Low & Low & High & Adaptive \\
        Optimal \( \rho \) & 1.0 (fixed) & 1.0 (fixed) & N/A & Dynamic (\( \rho \)) \\
        \bottomrule
    \end{tabular}
\end{table}
\section{Conclusion}
This paper presented FACET, a unified framework that redefines the role of feedback in energy-constrained IoT systems by integrating adaptive feedback channel coding with wireless power transfer. By exploiting the saturation effect inherent to feedback coding, FACET introduced a dual-purpose feedback mechanism that simultaneously enhances uplink communication efficiency and replenishes device energy. We formulated a fairness-aware min-max optimization problem to balance this tradeoff and developed an efficient algorithm leveraging alternating optimization and Lagrangian duality, with each subproblem admitting closed-form solutions.
Simulations showed that FACET achieves significant energy savings, extending device lifetime by up to $1.83\times$ and $13.4\times$ compared to feedback-only and WPT-only baselines, respectively. 
These findings establish FACET as a promising paradigm for next-generation green IoT networks.
\bibliography{reference}

\end{document}